\documentstyle[sprocl]{article}

\def\bea{\begin{eqnarray}}
\def\eea{\end{eqnarray}}
\def\beq{\begin{equation}}
\def\eeq{\end{equation}}

\def\ga{\gamma}
\def\de{\delta}

\def\si{\sigma}

\def\ps{\psi}
\def\om{\omega}

\def\De{\Delta}

\def\fr#1#2{{{#1} \over {#2}}}

\def\half{{\textstyle{1\over 2}}}
\def\frac#1#2{{\textstyle{{#1}\over {#2}}}}

\def\lsim{\mathrel{\rlap{\lower4pt\hbox{\hskip1pt$\sim$}}
    \raise1pt\hbox{$<$}}}
\def\gsim{\mathrel{\rlap{\lower4pt\hbox{\hskip1pt$\sim$}}
    \raise1pt\hbox{$>$}}}
\def\sqr#1#2{{\vcenter{\vbox{\hrule height.#2pt
         \hbox{\vrule width.#2pt height#1pt \kern#1pt
         \vrule width.#2pt}
         \hrule height.#2pt}}}}

\begin{document}
\setcounter{page}{103}

\title{THEORETICAL ANALYSIS OF CPT AND \\
LORENTZ TESTS IN PENNING TRAPS}

\author{R. BLUHM}

\address{Physics Department, Colby College,
Waterville, ME 04901, USA\\E-mail: rtbluhm@colby.edu}

%%%%%%%%%%%%%%%%%%%%%%%%%%%%%%%%%%%%%%%%%%%%%%%%%%%%%%%%%%%%%%
% You may repeat \author \address as often as necessary      %
%%%%%%%%%%%%%%%%%%%%%%%%%%%%%%%%%%%%%%%%%%%%%%%%%%%%%%%%%%%%%%

\maketitle\abstracts{ 
The CPT theorem has been tested 
to very high precision
in a variety of experiments involving 
particles and antiparticles confined within Penning traps.
These tests include comparisons of anomalous magnetic moments
and charge-to-mass ratios of electrons and positrons,
protons and antiprotons,
and hydrogen ions and antiprotons. 
We present a theoretical analysis of possible signals for CPT
and Lorentz violation in these systems.
We use the framework of Colladay and Kosteleck\'y,
which consists of a general extension of the
$SU(3) \times SU(2) \times U(1)$ standard model
including possible CPT and Lorentz violations
arising from spontaneous symmetry breaking at a 
fundamental level, 
such as in string theory.
We work in the context of an extension of quantum
electrodynamics to examine CPT and Lorentz tests
in Penning traps.
Our analysis permits a detailed study of the effectiveness
of experimental tests of CPT and Lorentz symmetry
performed in Penning traps.
We describe possible signals that might appear in principle,
and estimate bounds on CPT and Lorentz violation
attainable in present and future experiments.
}

\section{Introduction}

The CPT theorem \cite{cpt}
and Lorentz symmetry have both been
tested to very high accuracy in a variety of 
physical systems.\cite{pdg}
Papers presented at this meeting have described experiments in
astrophysical, nuclear, particle, and atomic systems,
all of which provide very stringent bounds on possible
CPT or Lorentz breaking.
To date,
the best bound on CPT
has been obtained in particle-physics
experiments involving neutral kaons.
Since different particle sectors are largely independent,
it is important to consider possible CPT and Lorentz
breaking in all particle sectors,
including mesons, leptons, baryons, and gauge bosons.
While kaon experiments clearly 
provide the best test of CPT in the meson sector,
it is interesting to note that
the sharpest tests of CPT breaking in both the lepton and 
baryon systems have not been obtained in high-energy
particle experiments.
Instead,
low-energy experiments
on single isolated particles in Penning traps
have yielded the best bounds on CPT
in the lepton and baryon sectors.
These experiments involve comparisons of electrons and positrons,
protons and antiprotons,
and hydrogen ions and antiprotons.

One consequence of CPT invariance is that
particles and antiparticles have equal 
charge-to-mass ratios and gyromagnetic ratios. 
Experiments in Penning traps are ideally suited
for making very precise comparisons of these quantities.
A Penning trap
\cite{penning}
captures a single charged particle in the cavity
between two cap electrodes and a ring electrode.
The electrodes are charged and create a quadrupole electric field.
A static magnetic field is created using external current coils.
A charged particle in the trap is bound due to the combination of
the static electric and magnetic fields.
By nesting two Penning traps,
particles and antiparticles can be probed and quickly switched
in the same magnetic field,
but with the electric field reversed.
The dominant structure of the energy levels 
for spin-$\half$ particles at low temperature is that
of relativistic Landau levels,
with two ladders of energies for the two spin states.
Transition frequencies between these levels can be measured with
very high precision.

Typically,
two types of frequency comparisons 
of particles and antiparticles are possible in Penning traps.
They involve making accurate measurements of the cyclotron
frequency $\om_c$ 
(for transitions between Landua levels with no spin flip)
and the anomaly frequency $\om_a$ 
(for transitions between Landua levels accompanied by a spin flip)
of single isolated particles confined in the trap.
The first type of experiment is an anomalous magnetic
moment or $g-2$ experiment.
These compare the ratio $2 \om_a / \om_c$ for particles
and antiparticles.
In the context of conventional 
quantum electrodynamics,
this ratio equals $g-2$ for the particle or antiparticle.
The second type of experiment compares values of $\om_c \sim q/m$,
where $q>0$ is the magnitude of the charge and $m$ is the mass. 
These therefore involve comparisons of the charge-to-mass ratios
for the particle and antiparticle.

Both $g-2$ and charge-to-mass ratio experiments have
been performed with electrons and positrons.
With protons and antiprotons,
however,
only charge-to-mass ratio comparisons have been performed.
Because the magnetic moments of protons and antiprotons
are much weaker than those of electrons and positrons,
$g-2$ experiments with protons and antiprotons
require much lower temperatures and greater sensitivity for
detecting spin-flip transitions.
Although these $g-2$ experiments with protons and antiprotons
have not been performed to date
some suggestions for making these experiments feasible 
in the future exist in the
literature.\cite{hw,qg}

To compare the sensitivities of CPT tests in Penning traps
with those in the meson system,
we list some of the relevant figures of merit.
The conventional figure of merit in the neutral kaon system
is given by
\beq
r_K \equiv \fr {|m_K - m_{\overline{K}}|} {m_K}
\lsim 2 \times 10^{-18}
\quad ,
\label{rK}
\eeq
whereas for $g-2$ experiments
with electrons and positrons,\cite{vd}
electron-positron charge-to-mass ratio experiments,\cite{schwin81}
and charge-to-mass ratio experiments with
protons and antiprotons,\cite{gg1,gg2}
respectively,
the conventional figures of merit are given as
\beq
r_g^e \equiv 
\fr {|g_{e^-} - g_{e^+}|} {g_{\rm avg}}
\lsim 2 \times 10^{-12}
\quad ,
\label{rg}
\eeq
\beq
r_{q/m}^e \equiv 
\fr {\left|({q_{e^-}}/{m_{e^-}})
- ({q_{e^+}}/{m_{e^+}})\right|} {|q/m|_{\rm avg}}
\lsim 1.3 \times 10^{-7}
\quad ,
\label{rqme}
\eeq
\beq
r_{q/m}^p \equiv \fr {|(q_p/m_p)
- (q_{\overline{p}}/m_{\overline{p}})|} {|q/m|_{\rm avg}}
\lsim
1.5 \times 10^{-9}
\quad .
\label{rqmp}
\eeq
Recently,
an experiment comparing the cyclotron frequencies of
hydrogen ions $H^-$ and antiprotons has been performed
in a Penning trap.\cite{gg2}
This experiment has the advantage that both particles 
in the trap have
the same electric charge,
thereby reducing systematic errors associated with 
reversing the sign of the electric field.
An improved charge-to-mass ratio comparison 
for protons and antiprotons has been
obtained from these results, 
which is given by
\beq
r_{q/m}^{H^-} \equiv \fr {|(q_p/m_p)
- (q_{\overline{p}}/m_{\overline{p}})|} {|q/m|_{\rm avg}}
\lsim
9 \times 10^{-11}
\quad .
\label{rqmH}
\eeq

Measurements of frequencies in Penning traps typically have
parts-per-billion (ppb) accuracies,
which are four or five orders of magnitude
better than the measurements made in kaon experiments.
This raises some interesting questions concerning the
sensitivity of these experiments to different possible
types of CPT breaking.
One goal of this work is to understand the Penning-trap experiments
better and to address the question of why
they do not provide sharper tests of CPT.
To accomplish this,
we must work in the context of a theoretical framework
that permits CPT breaking.
Such a framework has been developed
by Colladay and Kosteleck\'y.\cite{ck}

In the following sections,
the parts of the framework providing an extension
of quantum electrodynamics are described,
and the results of our theoretical 
analysis of CPT and Lorentz
tests in Penning traps are presented.
In particular,
we analyze
$g-2$ experiments on electrons and positrons,\cite{bkr1}
charge-to-mass ratio experiments on protons and antiprotons,
and comparisons of cyclotron frequencies for $H^-$ and
antiprotons.\cite{bkr2}
Since the framework we use includes both a CPT-violating sector
and a CPT-preserving sector
(both of which violate Lorentz symmetry)
in addition to investigating the sensitivity of Penning-trap
experiments to CPT,
we also examine how these experiments test CPT-preserving
Lorentz symmetry.

\section{Theoretical Framework}

The theoretical framework of Colladay and Kosteleck\'y \cite{ck}
is an extension of 
the $SU(3) \times SU(2) \times U(1)$ standard model. 
It originates from the idea of spontaneous CPT and 
Lorentz breaking
in a more fundamental theory.\cite{kps}
This type of CPT violation is a possibility in string
theory because the usual axioms of the CPT theorem do
not apply to extended objects like strings.
In a theory with spontaneous symmetry breaking,
the dynamics of the action remains CPT invariant,
which means the framework can preserve desirable features
of quantum field theory
such as gauge invariance, power-counting renormalizability,
and microcausality.
CPT and Lorentz violation occurs only in the
solutions of the equations of motion.
This mechanism is similar to the spontaneous breaking of the 
electroweak theory in the standard model.

In our analysis of Penning-trap experiments,
we use a restriction of the full
particle-physics framework to quantum
electrodynamics.
The effects of possible CPT and Lorentz violation
in this context
lead to a modification of the Dirac equation.
The modified form (in units with $\hbar = c = 1$) 
is given by
$$
( i \ga^\mu D_\mu - m - a_\mu \ga^\mu
- b_\mu \ga_5 \ga^\mu - \half H_{\mu \nu} \si^{\mu \nu}
\quad\quad\quad\quad\quad
$$
\beq 
\quad\quad\quad\quad\quad
+ i c_{\mu \nu} \ga^\mu D^\nu 
+ i d_{\mu \nu} \ga_5 \ga^\mu D^\nu) \ps = 0
\quad .
\label{dirac}
\eeq
Here, 
$\ps$ is a four-component spinor,
$A_\mu$ is the electromagnetic field,
$i D_\mu \equiv i \partial_\mu - q A_\mu$
is the covariant derivative,
and $a_\mu$, $b_\mu$, 
$H_{\mu \nu}$, $c_{\mu \nu}$, $d_{\mu \nu}$
are the parameters describing possible violations
of CPT and Lorentz symmetry.
The terms involving
$a_\mu$, $b_\mu$ 
break CPT
and those involving 
$H_{\mu \nu}$, $c_{\mu \nu}$, $d_{\mu \nu}$ 
preserve CPT,
while all five terms break Lorentz symmetry.

Since no CPT or Lorentz breaking has been observed 
in experiments to date,
the quantities $a_\mu$, $b_\mu$, 
$H_{\mu \nu}$, $c_{\mu \nu}$, $d_{\mu \nu}$ 
must all be small.
We can estimate the suppression scale for these
quantities by taking the scale governing the
fundamental theory as the Planck mass $m_{\rm Pl}$
and the low-energy scale as the electroweak 
mass scale $m_{\rm ew}$.
The natural suppression scale for Planck-scale effects
in the standard model would then be of order
$m_{\rm ew}/m_{\rm Pl} \simeq 3 \times 10^{-17}$.
If instead,
we consider the electron mass scale as the low-energy scale,
we obtain 
$m_{\rm e}/m_{\rm Pl} \simeq 5 \times 10^{-23}$.
Since a more fundamental theory (which would determine these
parameters more precisely) remains unknown,
these ratios give only an approximate indication of
the suppression scale.

We use this theoretical framework to analyze comparative tests of
CPT and Lorentz symmetry on particles and antiparticles
in Penning traps.
Some technical issues include the following.
First,
the time-derivative couplings in
Eq.\ \ref{dirac} alter the standard procedure for obtaining
a hermitian quantum-mechanical hamiltonian operator.
To overcome this,
we perform a field redefinition at the lagrangian
level that eliminates the additional time derivatives.
Second,
to obtain a hamiltonian for the antiparticle,
we use charge conjugation to find the Dirac equation
describing the antiparticle.
Perturbative calculations can then be carried out for
both the particle and antiparticle,
and the leading-order effects of CPT and Lorentz
breaking can be obtained.

\section{ Electron-Positron Experiments}

Experiments testing CPT in the electron-positron
system compare cyclotron frequencies
$\om_c$ and anomaly frequencies $\om_a$ 
of particles and antiparticles in a Penning trap.
A result of the CPT theorem is that
electrons and positrons of opposite spin in a
Penning trap with the
same magnetic fields but opposite electric fields should
have equal energies.
The experimental relations $g-2 = 2 \om_a / \om_c$ and 
$\om_c = qB/m$ provide connections to the quantities
$g$ and $q/m$ that appear in the figures of merit
$r_g^e$ and $r_{q/m}^e$.
Calculations are performed using 
Eq.\ \ref{dirac} to obtain possible shifts in the
energy levels due to either CPT-breaking or CPT-preserving
Lorentz violation.
The effectiveness of Penning-trap
experiments on electrons and positrons
as tests of both CPT-breaking and
CPT-preserving Lorentz violation
can then be analyzed.
From the calculated energy shifts we determine how the
frequencies $\om_c$ and $\om_a$ are affected and whether
the conventional figures of merit are appropriate.

The dominant contributions to the energy 
of an electron or positron in a Penning trap come from
interactions with the
constant magnetic field of the trap.
Interactions with the quadrupole electric field generate smaller effects.
In a perturbative treatment,
the dominant CPT- and Lorentz-breaking effects can therefore
be obtained by working with the relativistic Landau levels
as unperturbed states.
Conventional perturbations,
such as the usual corrections to the anomalous magnetic moment,
do not break CPT or Lorentz symmetry and
are the same for electrons and positrons.
Any violations of CPT or Lorentz symmetry result in
either differences between electrons and positrons
or in unconventional effects
such as diurnal variations in measured frequencies.

%\begin{figure}[t]
%\psfig{figure=fig1.ps,height=4.0in,width=6.0in}
%\caption{\label{fig:ab}}
%\end{figure}

Our calculations 
\cite{bkr1}
show that the leading-order corrections to
the energies $E_{n,s}^{e^-}$ for the electron  
and $E_{n,s}^{e^+}$ for the positron 
due to the effects of CPT and Lorentz violation are
$$
\quad
\de E_{n,\pm 1}^{e^-} \approx a_0^e \mp b_3^e
- c_{00}^e m_e \pm d_{30}^e m_e \pm H_{12}^e
$$
\beq
\quad\quad\quad\quad\quad\quad
- \half (c_{00}^e + c_{11}^e +c_{22}^e) (2n + 1 \pm 1) \om_c 
\quad .
\label{Eelec}
\eeq
$$
\quad
\de E_{n,\pm 1}^{e^+} \approx - a_0^e \mp b_3^e
- c_{00}^e m_e \mp d_{30}^e m_e \mp H_{12}^e
$$
\beq
\quad\quad\quad\quad\quad\quad
- \half (c_{00}^e + c_{11}^e + c_{22}^e) (2n + 1 \mp 1) \om_c 
\quad .
\label{Epos}
\eeq
From these we find the modified transition frequencies including
the leading-order effects of CPT and Lorentz breaking.
These are given by
\beq
\om_c^{e^-} \approx \om_c^{e^+} \approx
(1 - c_{00}^e - c_{11}^e - c_{22}^e) \om_c
\quad ,
\label{wcelec}
\eeq
\beq
\om_a^{e^\mp} \approx \om_a
\mp 2 b_3^e + 2 d_{30}^e m_e + 2 H_{12}^e
\quad .
\label{waelec}
\eeq
Here,
$\om_c$ and $\om_a$ represent the
unperturbed electron or positron frequencies,
while $\om_c^{e^\mp}$ and $\om_a^{e^\mp}$ denote
the frequencies including corrections.
Superscripts have been added to the parameters $b_\mu$, etc.\
to denote that they describe the electron-positron system.

From these relations we find the electron-positron differences
for the cyclotron and anomaly frequencies to be
\beq
\De \om_c^e \equiv \om_c^{e^-} - \om_c^{e^+} \approx 0
\quad , 
\label{delwc}
\eeq
\beq 
\De \om_a^e  \equiv \om_a^{e^-} - \om_a^{e^+} \approx - 4 b_3^e
\quad .
\label{delwa}
\eeq
In the context of this framework, 
comparisons of
cyclotron frequencies to leading order
do not provide a signal for CPT or Lorentz breaking,
since the corrections to $\om_c$ for electrons and
positrons are equal.
On the other hand,
comparisons of $\om_a$ provide unambiguous tests of CPT.

We also find that there are no leading-order
corrections due to CPT or Lorentz violation
to the $g$ factors for either
electrons or positrons.
This leads to some unexpected results
concerning the figure of merit $r_g$ in
Eq.\ \ref{rg}.
With $g_{e^-}$ and $g_{e^+}$ equal to leading order,
we find that $r_g$ vanishes,
which would seem to indicate the absence of CPT breaking.
However,
this conclusion would be incorrect
because the framework we are working 
in contains explicit CPT violation.
In addition,
calculations in the context of our framework
show that with $\vec b \ne 0$ the
experimental ratio $2 \om_a / \om_c$ depends on
the magnetic field and 
is undefined in the limit of a vanishing $B$ field.
Because of this,
the usual relation $g-2 = 2 \om_a / \om_c$ does not hold
in the presence of CPT violation.
For these reasons,
we conclude that in the context of our framework
the figure of merit $r_g$ in
Eq.\ \ref{rg} is inappropriate,
and an alternative is suggested next.

Since it is a prediction of the CPT theorem that 
electron and positron states of opposite spin 
in the same magnetic field have equal energies,
we propose as a model-independent figure of merit
\beq
r^e_{\om_a}
\equiv \fr{|{E}_{n,s}^{e^-} - {E}_{n,-s}^{e^+}|}
{{E}_{n,s}^{e^-} }
\quad .
\label{re}
\eeq
Here, ${E}_{n,s}^{e^\mp}$ are the Landau-level energies,
with $n$ denoting the Landau level,
and $s=\pm 1$ the spin.
In the context of our framework,
we find $r^e_{\om_a} \approx |2 b_3^e | / m_e$,
which can be bounded by experiments.
Assuming ppb frequency resolutions,
we estimate as a bound,
\beq
r^e_{\om_a} \lsim 10^{-21}
\quad .
\label{relim}
\eeq

The figure of merit $r^e_{\om_a}$
is compatible with the corresponding
figure of merit $r_K$ 
which describes the neutral-kaon system.
This is because both figures of merit involve energy ratios,
which makes comparisons across experiments more meaningful.
In contrast,
the figures of merit $r_g^e$ and $r_K$
involve ratios of different physical quantities.
Moreover,
our estimated bound for $r^e_{\om_a}$ 
is more in line with the high precision 
that is experimentally accessible in frequency
measurements in a Penning trap
and appears to improve on the bound given in
terms of $r_K$.
It is important to stress,
however, that performing CPT tests in the 
meson sector remains essential
because CPT violation in this sector is controlled 
by distinct CPT-violating parameters that appear 
only in the quark sector.\cite{k98}

Alternative signatures of CPT 
and Lorentz violation can be
considered as well.\cite{bkr2}
These include possible diurnal variations in the anomaly
and cyclotron frequencies.
We estimate bounds for these quantities
based on ppb accuracies in $\om_a$ and $\om_c$.
They are
\beq
r^e_{\om_{a}^{\mp},\rm diurnal} 
\approx
\fr {2|\mp b_3^e + d_{30}^e m_e + H_{12}^e|}{m_e}
\lsim 10^{-21}
\quad ,
\label{rwadiurnal}
\eeq
\beq
r^e_{\om_c, \rm diurnal} \approx 
\fr {|c_{11}^e + c_{22}^e| \om_c} {m_e} \lsim 10^{-18}
\quad .
\label{rwcdiurnal}
\eeq
Tests for these effects would provide 
bounds on some of the components of the
CPT-preserving but Lorentz-violating parameters
$c_{\mu \nu}^e$, $d_{\mu \nu}^e$, and $H_{\mu \nu}^e$.

One type of experiment searching for diurnal variations 
would involve the electron alone or the positron alone
in a Penning trap.
Diurnal variations in the cyclotron and
anomaly frequencies would occur because
the spatial components of the parameters in 
Eq.\ref{waelec}
would change as the Earth rotates.
A figure of merit can be defined which is based on
the relative size of the diurnal energy variations.
First,
consider the following quantities for the
electron and positron:
\beq
\De^{e}_{\om_a^{e^-}} 
\equiv 
\fr {|{E}_{0,+1}^{e^-} - {E}_{1,- 1}^{e^-}|}
{{E}_{0,-1}^{e^-}} 
\quad ,  \quad \quad 
\De^{e}_{\om_a^{e^+}} 
\equiv 
\fr {|{E}_{0,-1}^{e^+} - {E}_{1,+1}^{e^+}|}
{{E}_{0,+1}^{e^+}} 
\quad .
\label{Deleorpdnlom}
\eeq
Suitable figures of merit $r^e_{\om_{a}^-,\rm diurnal}$  
and $r^e_{\om_{a}^+,\rm diurnal}$  
can then be defined as 
the amplitude of the diurnal variations 
in $\De^{e}_{\om_a^{e^-}}$
and $\De^{e}_{\om_a^{e^+}}$, 
respectively.
In the context of the framework we are using,
we compute that
\beq
r^e_{\om_{a}^{\mp},\rm diurnal} 
\approx
\fr {2|\mp b_3^e + d_{30}^e m_e + H_{12}^e|}{m_e}
\quad .
\eeq
Among the experimental issues involved in obtaining a
bound on $r^e_{\om_{a}^{\mp},\rm diurnal}$ is 
maintaining stability in the magnetic field.
For example,
drifts in the magnetic field at a 
level of about 5 parts in $10^9$ over the duration of the experiment
would correspond to a 1 Hz frequency resolution.
The data would then need to be plotted and fitted
as a function of the orientation of the magnetic field 
with respect to a celestial coordinate system.

Bounds obtained in an experiment on electrons alone
or positrons alone would involve the combination 
$\mp b_3^e + d_{30}^e m_e + H_{12}^e$
of parameters in the standard-model extension.
The dominant signal would therefore involve
corrections to the anomaly and cyclotron frequencies 
which exhibit periodicities of approximately 24 hours.
Subleading order corrections might exhibit 12-hour periodicities.
However, these effects would be suppressed relative
to the leading-order effects.
All three of the quantities
$b_3^e$, $d_{30}^e$, and $H_{12}^e$ 
break Lorentz symmetry,
but only the coupling $b_3^e$ breaks CPT.
If a signal were detected,
it would indicate Lorentz breaking
but not necessarily CPT violation.
A subsequent experiment 
comparing anomaly frequencies of electrons and positrons
which would bound the CPT-breaking parameter $b_3^e$
in isolation would then need to be performed.

A preliminary analysis of this type of experiment on electrons 
alone has recently been performed.\cite{mit}
With a precision of approximately 1 Hz
in detecting diurnal variations, 
an estimated bound on Lorentz breaking is given as
\beq
r^e_{\om_{a}^{\mp},\rm diurnal} \lsim 10^{-20}
\quad .
\eeq

\section{Proton-Antiproton Experiments}

We also investigate the sensitivity to CPT and Lorentz violations
of charge-to-mass-ratio experiments and
possible future $g-2$ experiments involving protons
and antiprotons in Penning traps.
In this analysis,
it suffices to work at the level of an effective theory
in which the protons and antiprotons are regarded as
basic objects described by a Dirac equation.
The coefficients $a_\mu^p$, $b_\mu^p$, $H_{\mu \nu}^p$, 
$c_{\mu \nu}^p$, $d_{\mu \nu}^p$ represent effective
parameters,
which at a more fundamental level depend on underlying
quark interactions.

To leading order,
we find the proton-antiproton differences
for the cyclotron and anomaly frequencies are
\beq
\De \om_c^p \equiv \om_c^{p} - \om_c^{\bar p} \simeq 0
\quad , 
\label{omcp}
\eeq
\beq
\De \om_a^p  \equiv \om_a^{p} - \om_a^{\bar p} \simeq 4 b_3^p
\quad .
\label{omap}
\eeq
Assuming $\om_a^p$ and $\om_a^{\bar p}$ 
can be measured with ppb accuracies,
and defining an appropriate figure of merit,
we estimate for $g-2$ experiments
\beq
r^p_{\om_a}
\equiv \fr{|{ E}_{n,s}^{p} - { E}_{n,-s}^{\bar p}|} 
{{ E}_{n,s}^{p}}
\lsim 10^{-24}
\quad ,
\label{romap}
\eeq
whereas in experiments searching for diurnal variations
we estimate
\beq
r^p_{\om_{a}^{\mp},\rm diurnal} 
\approx
\fr {2|\mp b_3^p + d_{30}^p m_p + H_{12}^p|}{m_p}
\lsim 10^{-24}
\quad ,
\label{pwadiurnal}
\eeq
\beq
r^p_{\om_c, \rm diurnal} \approx 
\fr {|c_{11}^p + c_{22}^p| \om_c} {m_p} \lsim 10^{-24}
\quad .
\label{pwcdiurnal}
\eeq

A recent experiment\cite{gg2} 
compares antiproton cyclotron frequencies
with those of an $H^-$ ion instead of a proton.
This comparison provides
a sharp test of CPT-preserving Lorentz symmetry.
In the context of our frameowrk,
the difference 
between the cyclotron frequencies of the $H^-$ hydrogen ion and 
the antiproton can be computed and is given by 
$$
\De \om_{c,\rm th}^{H^-}
\approx
(c_{00}^{p} + c_{11}^{p} +c_{22}^{p}) 
( \om_c - \om_c^{H^-} )
\quad\quad\quad\quad
$$
\beq
\quad\quad\quad\quad
- \fr{2m_e}{m_p} 
(c_{00}^{e} + c_{11}^{e} + c_{22}^{e}
-c_{00}^{p} - c_{11}^{p} -c_{22}^{p}) 
\om_{c}^{H^-} 
\quad .
\label{romacH}
\eeq
The estimated bound that follows from this is\cite{bkr2}
\beq
r^{H^-}_{\om_c} \approx
|\De \om_{c,\rm th}^{H^-}| / {m_p} 
\lsim 10^{-25}
\quad .
\label{rH}
\eeq

\section{Conclusions}

In summary,
we find that the use of a general theoretical framework
incorporating CPT and Lorentz violation permits a detailed
investigation of possible experimental signatures in
Penning-trap experiments.

In the electron-positron system,
our results indicate that the sharpest tests of CPT
in Penning-trap experiments emerge from comparisons of
anomaly frequencies in $g-2$ experiments
and that bounds of order $10^{-21}$ are attainable.
In the context of our theoretical framework,
we find that
the conventional figure of merit $r_g^e$ does not provides an
appropriate bound on CPT,
and we have suggested an alternative.
We find that comparisons of cyclotron frequencies are not 
sensitive to leading-order CPT or Lorentz violation,
whereas diurnal variations in $\om_a$ and $\om_c$ can provide
new signals for Lorentz violation with bounds
of order $10^{-21}$ and $10^{-18}$,
respectively.
Experiments searching for 
diurnal variations in electrons
alone can provide a bound on Lorentz breaking at a level
of approximately $10^{-20}$.

In the proton-antiproton system,
our results show that future $g-2$ experiments 
on protons and antiprotons could
provide stringent test of CPT,
with bounds of order $10^{-24}$.
Experiments searching for 
diurnal variations in the proton-antiproton system
can also provide bounds on Lorentz and CPT breaking at a level
of approximately $10^{-24}$.
A recent comparison of $H^-$ and antiproton cyclotron frequencies
have provided a new test of CPT-preserving Lorentz invariance at a
level of $10^{-25}$.

Table I contains a summary of the estimated bounds attainable
in Penning-trap experiments in the three systems considered here.

\begin{table}[h]
\caption{Estimated CPT- and Lorentz-violating bounds 
for electron-postron, proton-antiproton, 
and $H^-$-antiproton experiments in Penning traps.
The estimated bounds are based on 
ppb accuracies in $\om_a$ and $\om_c$ 
(except in the $H^-$ $\bar p$ experiments
which have $\sim 10^{-10}$ accuracy).
The first two columns specify the type of experiment.
The third column lists the figures of merit,
while the fourth gives the corresponding bounds
estimated from current or future experiments.
The fifth column shows the affected parameters,
and the last shows which symmetry is tested,
CPT-violating Lorentz symmetry (CPT) or 
CPT-preserving Lorentz symmetry (Lorentz).\label{tab:exp}}
\vspace{0.2cm}
\begin{center}
%\footnotesize
\begin{tabular}{|l|c|c|c|l|l|}
\hline
System & Expt.\ & Fig.\ Merit & Est.\ Bound & Parms.\ &Test
\\
\hline\hline
$e^- \, e^+$ & $\De \om_a$ & $r^e_{\om_a}$ & $10^{-21}$ & 
   $b_j^e$ & CPT \\[2mm] 
\cline{2-6}
 & $\om_a$  \, diurnal  & $r^e_{\om_a, \rm diurnal}$
 & $10^{-21}$ & $d_{j0}^e,\  H_{jk}^e$ & Lorentz \\[2mm]
\cline{2-6}
 & $\om_c$ \, diurnal  & $r^e_{\om_c, \rm diurnal}$
 & $10^{-18}$ & $c_{jj}^e$ & Lorentz \\[2mm]
\cline{1-6}
$p \, \bar p$ & $\De \om_a$ & $r^p_{\om_a}$ & $10^{-24}$ & 
   $b_j^p$ & CPT \\[2mm] 
\cline{2-6}
 & $\om_a$  \, diurnal  & $r^e_{\om_a, \rm diurnal}$
 & $10^{-24}$ & $d_{j0}^p,\  H_{jk}^p$ & Lorentz \\[2mm]
\cline{2-6}
 & $\om_c$ \, diurnal  & $r^e_{\om_c, \rm diurnal}$
 & $10^{-24}$ & $c_{jj}^p$ & Lorentz \\[2mm]
\cline{1-6}
$H^- \, \bar p$ & $\De \om_c$ & $r^{H^-}_{\om_c}$
 & $10^{-25}$ & $c_{jj}^e$, $c_{jj}^p$ & Lorentz \\[2mm]
\hline
\end{tabular}
\end{center}
\end{table}

%\vfill\eject

\section*{Acknowledgments}
This work is supported in part by the National
Science Foundation under grant number PHY-9801869.

\end{document}